# Study of Thermal Expansion Coefficients of 2D Materials via Raman Micro-spectroscopy: Revisited

*Qianchi Feng,[1, ‡] Dongshan Wei,[2, ‡] Yudan Su,[1, ‡] Zhiguang Zhou,[1] Feng Wang[3,*] and Chuanshan Tian[1,*]*

[1]State Key Laboratory of Surface Physics and Key Laboratory of Micro- and Nano-Photonic Structures (MOE), Department of Physics, Fudan University, Shanghai 200433, China

[2]School of Electronic Engineering and Intelligentization, Dongguan University of Technology, Dongguan, Guangdong, 523808, China

[3]Department of Chemistry and Biochemistry, University of Arkansas, AR 72703

ABSTRACT: We report a joint study, using Raman micro-spectroscopy and molecular dynamics simulations, on the substrate effect on thermal properties of 2D materials and revisit measurement of thermal expansion coefficient (TEC) of supported 2D film. Graphene is employed as a representative. We find that the out-of-plane coupling between graphene and substrate strongly affects the temperature-dependent vibrational modes and TEC of graphene. Density of states for long-wavelength out-of-plane oscillations is significantly reduced when graphene is supported on an alkane substrate. To account for the contribution of the out-of-plane coupling to TEC, a Raman micro-spectroscopic scheme is developed. The TEC of graphene on octadecyltrichlorosilane substrate is found to be $(-0.6\pm0.5)\times10^{-6}$/K at room temperature, which is fundamentally smaller than that of free-standing graphene. Our results shed light on the understanding of the interaction between 2D material and substrate, and offer a general recipe for optical measurement of TEC of a supported 2D film.

Two-dimensional (2D) materials, such as graphene, h-BN, silicene, h-MoS$_2$ and black phosphorus, have generated enormous interest because of its unique physical properties and promising applications in next-generation electronic and optoelectronic devices.[1-3] The thermal expansion coefficient (TEC, α) of a 2D material is a key physical property for both fundamental studies and practical applications. Due to their unique membrane structure, 2D materials may anomalously exhibit negative in-plane thermal expansion, which was firstly pointed out as "membrane effect" by Liftshitz,[4] attracting much attention in recent studies.

Theoretically, TEC of 2D materials can be calculated using quasi-harmonic approximations as shown in the following equation:[5]

$$\alpha = \frac{1}{a_0^2 \left.\frac{\partial^2 E}{\partial a^2}\right|_0} \sum_{k,s} c_v(k,s) \gamma(k,s) \tag{1}$$

Here, index "0" means the ground state, a is the lattice constant, and $E$ is the total energy of the system, $c_v(k,s)$ is the contribution of point k in s sub-band to the heat capacity. $\gamma(k,s)$ is the so-called Grüneisen parameter, with the form of $\gamma(k,s) = \left.\frac{-a_0}{\omega_{0,k,s}} \frac{\partial \omega_{k,s}}{\partial a}\right|_0$. Obviously, knowledge of phonon vibrational modes is crucial to understand the origin of negative TEC of some 2D materials. In a simplified physical picture, unlike its bulk counterpart, 2D structure leads to the ease of out-of-plane deformations, while the lattice is less "stretched".[5] The strong out-of-plane vibrations of 2D materials result in a tendency of in-plane contraction, which is known as the membrane effect. According to recent studies, out-of-plane phonon modes (also called as ZA phonon mode, bending mode or flexural mode) of 2D materials had been shown to be responsible for the phenomenon of negative TEC.[5-13] When the 2D material is stretched laterally, the strain leads to an increase of ZA phonon frequencies in addition to larger in-plane lattice constants, which results in negative Grüneisen parameter and thus the negative $\alpha$ according to Eq. (1). Note that 2D materials are generally fabricated on a supporting substrate. The interaction between the 2D film and substrate not only introduces in-plane deformation in the film, but is expected to strongly suppress the out-of-plane modes and the membrane effect.[14, 15] Specifically, TEC of graphene was predicted to be less negative or even positive when substrate interaction is considered.[14, 15] However, previous experimental studies were mainly focused on in-plane stretch or compression of 2D materials induced by substrate.[17-19] Experimentally, the important role of the out-of-plane coupling and its effect on the thermal expansion remain unclear.

Several techniques had been applied to TEC measurements of 2D materials, such as the scanning electron microscope (SEM),[20] micromechanical resonator,[21] x-ray diffraction (XRD),[22] Fabry-Perrot (FP) interference,[23] digital image correlation (DIC),[24] atomic force microscopy (AFM),[13] transmission electron microscope (TEM)[25] and Raman spectroscopy[17-19, 26-28]. Unfortunately, the reported values of TEC of monolayer graphene diverge, varying from $-8.0\times10^{-6}$/K to $+1.6\times10^{-6}$/K at room temperature.[27] As discussed above, the out-of-plane coupling has crucial effect on TEC of 2D material. The neglect of proper consideration of such coupling in these studies could be one of the most important reasons for the discrepancies.[10, 14, 15] Moreover, we noticed that even the non-uniformity of in-plane strain can cause large inaccuracy in TEC measurements. For instance, previous Raman spectroscopic studies assumed uniform in-plane coupling and non-slip boundary conditions between sample and substrate when deducing TEC of graphene.[17-19, 27] But Graf *et al*. showed that the interaction between graphene and substrate can be far from uniform.[29] To obtain reliable TEC of a supported 2D material, it is of paramount importance to study the substrate effects by taking into account both the out-of-plane coupling and the non-uniformity of the in-plane strain.

In this letter, we report an *in situ* Raman micro-spectroscopic study on TEC of a 2D material on substrate. Graphene on octadecyltrichlorosilane(OTS)-coated glass was chosen as a representative. By introducing a tunable in-plane strain and monitoring the local Raman shifts of graphene as a function of temperature from 290 K to 390K, the contributions of in-plane and out-of-plane substrate effects were able to be separated. We found that, although the terminal methyl groups of OTS interact only weakly with graphene, resulting in weak in-plane strain on the graphene, their presence at close range still affects TEC of graphene appreciably. A Raman micro-spectroscopic scheme is developed to measure TEC of 2D materials with the out-of-plane coupling effect being considered. The experimentally deduced TEC of graphene on OTS was found to be $(-0.6\pm0.5)\times10^{-6}$/K, the amplitude of which is fundamentally smaller than that of free-standing graphene reported in literature.[5, 11, 20, 21, 23, 30] Our theoretical analysis shows the reduction of ZA phonon density of states at longer wavelength and confirms the vast reduction of contribution to TEC from the out-of-plane modes when graphene is supported on a slippery substrate with only out-of-plane coupling. These results improve our understanding on how crucial out-of-plane coupling is to TEC of a 2D film. The technique for TEC measurement used in this work paves the way for TEC study of other 2D materials on substrates.

Graphene samples used in this work were commercial CVD single-layer graphene grown on copper foil (6Carbon Technology Inc.). PMMA (molecular weight 95000) was used to transfer the graphene on to a piece of 170 μm-thick cover glass (VWR International, LLC) with or without OTS surface coating, respectively. Raman spectra of graphene G-, D- and 2D-bands were measured to confirm the number of graphene layers.[31] Temperature-dependent Raman spectra of graphene 2D-band were recorded from 290K to 390K. HeNe laser ($\lambda$ = 632.8 nm) was focused on the graphene sample to excite Raman signal though an long working distance objective (Olympus, 50×, 0.5 NA). The Raman microscope was used to monitor the spectral change of the sample at the same location with spatial resolution of 1.0 μm during temperature variation and tuning of sample strain.

The temperature-dependent Raman frequency shift of graphene on substrate is given by[17]

$$\Delta\omega(T) = \Delta\omega_g(T) + \Delta\omega_{sub}(T) \tag{2}$$

The first term, $\Delta\omega_g(T)$, is attributed to the combination of lattice expansion and anharmonic effect of graphene. The second term, $\Delta\omega_{sub}(T)$, originates from the substrate contribution, which consists of the lateral strain ($\Delta\omega_{ls}(T)$) and the out-of-plane coupling ($\Delta\omega_{oc}(T)$) contributions:

$$\Delta\omega_{sub}(T) = \Delta\omega_{oc}(T) + \Delta\omega_{ls}(T) \tag{3}$$

In particular, $\Delta\omega_{ls}(T)$ is induced by stress as a result of TEC mismatch between graphene and substrate as temperature increases, which can be expressed as[17]

$$\Delta\omega_{ls}(T) = \beta \cdot \varepsilon(T) = \beta \cdot \int_{297K}^{T} (\alpha_{sub}(\tau) - \alpha_{graphene}(\tau)) d\tau \tag{4}$$

Here, $T$ is the sample temperature, $\beta$ is the biaxial strain coefficient, $\varepsilon$ is the strain imposing on graphene. $\alpha_{sub}(\tau)$ and $\alpha_{graphene}(\tau)$ are TEC of substrate and graphene at temperature $\tau$, respectively. Considering the substrate-dependence and non-uniformity of the strain coefficient,[29, 32, 33] $\beta$ in Eq. (4) shall be replaced by $\beta_{eff}$ that must be measured at the same location throughout the measurement:

$$\Delta\omega_{ls}(T) = \beta_{eff} \cdot \varepsilon(T) = \beta_{eff} \cdot \int_{297K}^{T} (\alpha_{sub}(\tau) - \alpha_{graphene}(\tau)) d\tau \tag{5}$$

Experimentally, $\beta_{eff}$ can be determined by applying a known strain on the substrate at a certain temperature. As shown in Fig. 1(a), the graphene sample on cover glass was sandwiched between two curvature matched positive and negative spherical lenses to apply biaxial strain. With known thickness ($d$) of the substrate and radius of the lens curvature ($R$), $β_{eff}$ can be deduced by comparing the frequency shift of the Raman band ($\Delta\omega_{ls}$) due to the applied strain at a constant temperature.

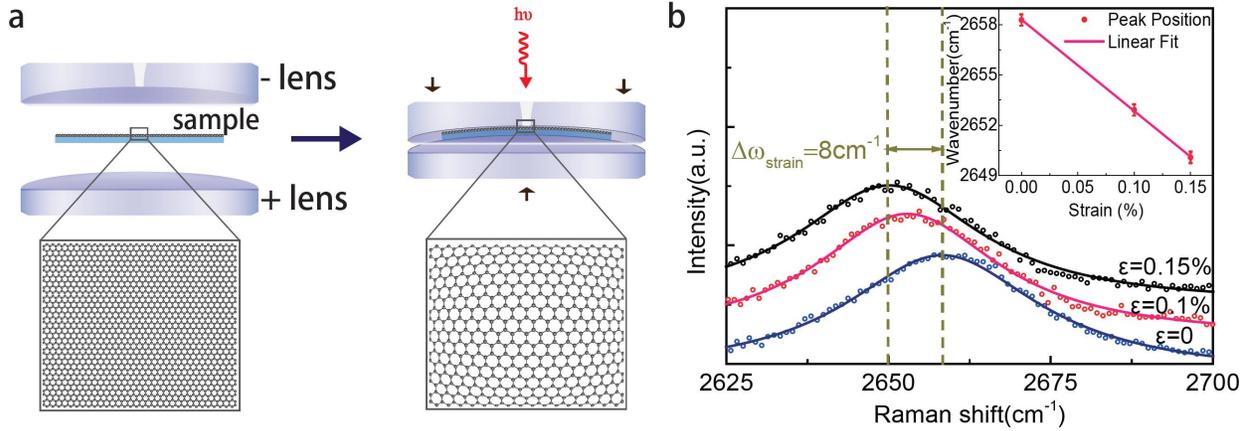

**Figure 1** (a) Experimental scheme to measure local $β_{eff}$ of a 2D film on substrate. The sample is bent by pressing the positive and negative spherical lenses with opposite curvature against one another. Biaxial strain is varied by choosing the lens pair with different curvature. A 1mm-diameter hole was drilled through the negative lens to avoid contact of the negative lens to graphene film. (b) The 2D band of graphene on cover glass under different biaxial strain. The vertical dashed lines indicate the Raman frequency shift between lateral strain $\varepsilon=0.15\%$ and $\varepsilon=0$. Inset: linear fit of peak positions of graphene 2D band as a function of strain.

Figure 1(b) shows a set of Raman spectra of graphene on bare glass at different strain imposed by the substrate. The 2D band of graphene shows linear dependence of frequency shift versus the applied lateral strain. The $β_{eff}$ is found to be 53cm$^{-1}$/%. It is worthwhile to mention that the value of $β_{eff}$ may vary from 35cm$^{-1}$/% to 64cm$^{-1}$/% at different locations on the same sample. This result signifies the importance of using microscope when studying the interaction between

2D film and substrate.

To demonstrate the importance of out-of-plane coupling between sample and substrate, we transferred graphene onto OTS-coated glass, where the lateral strain induced by substrate is expected to be weak.[34] Figure 2(a) presents a set of Raman spectra of graphene under substrate strain of 0%, 0.1% and 0.15%, respectively. The 2D-band is shifted merely by 1.5 cm$^{-1}$ when stretching the substrate by 0.15%, in contrast to 8.0 cm$^{-1}$ for graphene on glass (see Fig. 1(b)). The strain coefficient ($\beta_{eff}$) for graphene on OTS-coated glass is found to be 10 cm$^{-1}$/%. It indicates graphene is slipping on OTS surface when bending the substrate. As a result, according to Eq. (5), the lateral strain contribution to Raman shift is small when increasing the temperature because lateral expansion/contraction of graphene and the substrate is essentially decoupled. If the out-of-plane coupling between graphene and substrate were negligible as assumed in literature, the temperature-induced Raman shift ( $\Delta\omega\,(T_S)$ ) of graphene on OTS should have been close to that of free-standing graphene. Instead, as sketched in Fig. 2(b), the temperature-induced Raman shift of graphene on OTS is much larger than that of free-standing graphene when temperature increases from 290K to 390K. Even when the weak lateral strain contribution is taken into account (see supporting information for details), the resulting Raman shift versus temperature (orange curve in Fig. 2(b)) is still much smaller than the measured data. Such large difference signifies the important contribution of the out-of-plane coupling to the phonon vibrations of graphene.

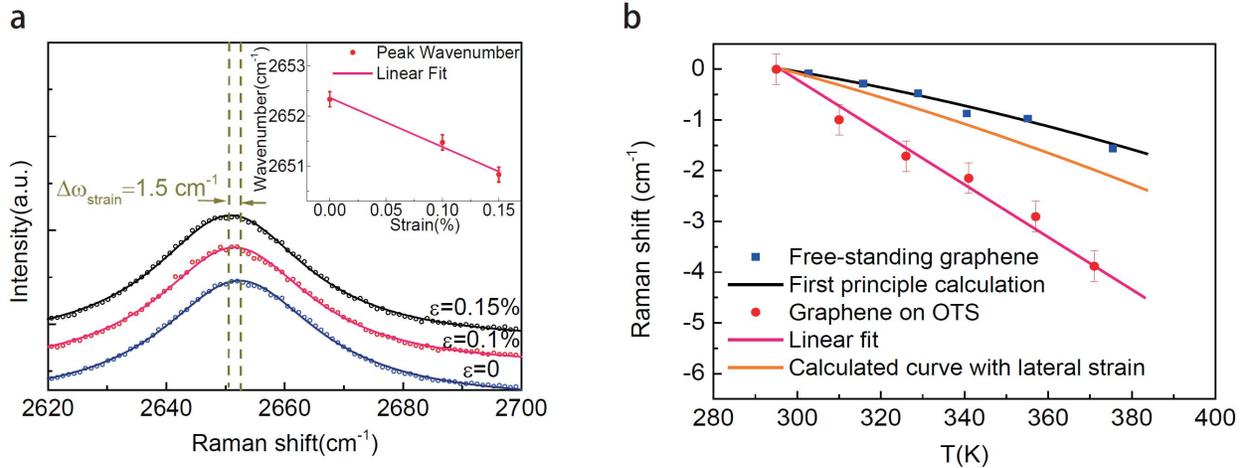

**Figure 2** (a) 2D Raman spectra of graphene on OTS under different biaxial strain; inset: linear fit of peak position of graphene 2D band versus strain. (b) Raman shift of graphene on OTS (red dot) in this work and free-standing graphene (blue dot) measured by Lin et al.[35] as a function of temperature. The black curve is the theoretical result by Bonini et al.[36] The orange curve is the calculated result assuming the interaction between graphene and substrate is merely lateral strain without the out-of-plane coupling.

To better elucidate the effect of out-of-plane coupling, molecular dynamics simulations were performed on graphene using the PPBE-G model[37, 38] with and without a model substrate. PPBE-G model was developed by force-matching density function theory with the PBE exchange correlation functional. The model has been shown to give good properties for free standing graphene. The substrate is modeled with a periodic array of $CH_4$ in a square lattice. The lattice spacing of $CH_4$ is chosen to be 4.5 Å mimicking the methyl group of OTS at similar areal density. The substrate is modeled using simple Lennard Jones potential from the OPLS-AA parameterization.[39] The perfectly flat substrate coupled with weak Lennard Jones interaction leads to a non-stick surface. The graphene is expected to be able to slide freely on the substrate resulting in a slippery support. Details of the computational modeling can be found in the supporting information.

Both free standing and supported graphene are expected to ripple, the amount of rippling can be quantified using the normal-normal correlation function,

$$G(R) = <\mathbf{n}(0) \cdot \mathbf{n}(R)> \tag{6}$$

Figure 3(a) compares the $G(R)$ of free and supported graphene. A perfectly flat graphene will have a $G(R)$ of one. Having a relatively large bending rigidity, graphene does not ripple much at room temperature. However, the small rippling is further diminished on the model substrate. The density of states of the out-of-plane vibrations (ZA modes) is best quantified by the Fourier transform of the height-height correlation function $\langle |h(\mathbf{q})|^2 \rangle$ presented in Fig. 3(b). It is interesting to note that the density of states at shorter wavelength (< 1nm) is not affected by the substrate. However, the substrate significantly reduces the density of states at longer wave-length. Such a suppression of longer wave-length bending is expected to affect both Raman peak shift and the TEC.

With the PPBE-G graphene model, the TEC of free standing graphene was computed to be $(-3.0 \pm 0.7) \times 10^{-6}$ K$^{-1}$ and that of supported graphene was $(2.0 \pm 0.7) \times 10^{-7}$ K$^{-1}$. The suppression of ZA vibrational modes normal to the substrate by the van der Waals interactions with the model substrate is thus strong enough to significantly reduce the in-plane contraction. This is not surprising since graphene can no longer ripple into the perpendicular dimension in presence of the out-of-plane coupling with the substrate. With the perfectly flat model substrate reducing the negative contribution to the TEC, the computed TEC reduced by more than a factor of ten in magnitude and becomes slightly positive.

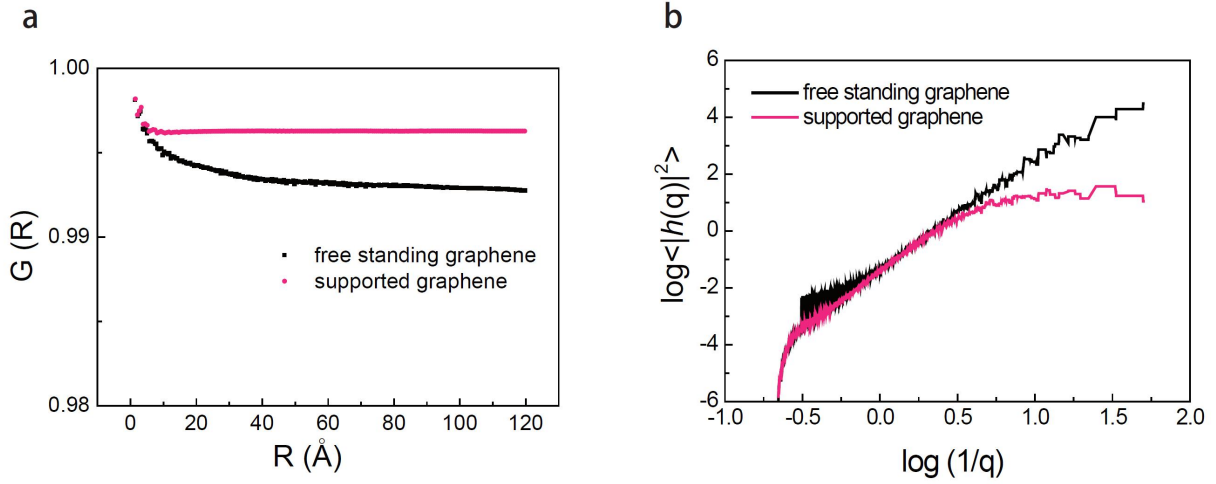

**Figure 3** (a) Normal-normal correlation function of free standing graphene and graphene on slippery array of $CH_4$ modeled as a square lattice. (b) The Fourier transform of the height-height correlation function shows the density of out-of-plane vibrational states as a function of wave length (1/q). Note that q has a unit of 1/Å.

The above experimental and theoretical results alert that it is problematic to ignore the effect of out-of-plane coupling in the measurement of graphene TEC as conducted in some previous experiments. In the following, we revisit the experimental determination of graphene TEC by taking into account the out-of-plane coupling. In order to obtain TEC using Raman spectroscopy, one needs to two sets of measurements with different lateral stain while the out-of-plane coupling and the $\Delta\omega_g(T)$ terms remains the same. As shown in Eq. (2) and (3), taking the difference between the two results gives the Raman shift solely induced by the lateral strain contribution. Note that the lateral strain imposing on graphene as temperature rises is caused by the difference of TEC between graphene and the substrate. With known TEC of the substrate, one can then deduce the TEC of graphene.

We therefore designed a patterned substrate consisting of 25μm diameter OTS disks surrounded by gold film on glass substrate as illustrated in Fig. 4(a), then transferred graphene on the top. The interaction between graphene and gold on glass is sufficiently strong such that a known strain can be applied to graphene during the expansion of the substrate.[40, 41] As temperature rises, graphene on the OTS domain is stretched by the surrounding gold-coated glass. After collecting a set of Raman spectra of graphene at a marked location on the OTS domain versus temperature, we used tightly focused femtosecond laser pulses to cut graphene along the edge of the OTS disk. It allows isolation of the piece of graphene on the OTS disk from the rest as sketched in Fig. 4(b). Then, another set of Raman spectra of the now isolated graphene were recorded versus temperature. In the above-mentioned two sets of measurements, $\Delta\omega_g(T)$ and the out-of-plane coupling contribution ($\Delta\omega_{oc}(T)$) in Eq. (2) and (3) are identical because the same region of graphene on the OTS disk was probed. Thus, the lateral strain term ($\Delta\omega_{ls}(T)$) can be obtained by taking the differences of the two sets of Raman shifts:

$$\Delta\omega_{ls}^{pat}(T) - \Delta\omega_{ls}^{iso}(T) = (\beta_{eff}^{pat} - \beta_{eff}^{iso})\int_{297K}^{T}(\alpha_{sub}(\tau) - \alpha_{graphene}(\tau))d\tau \tag{7}$$

Here, the superscript "*pat*" and "*iso*" denotes the corresponding parameters of patterned and isolated sample, respectively. With known $\alpha_{sub}(T)$, and $\beta_{eff}$ calibrated *in situ* following the protocol presented in Fig. 1(a), TEC of graphene ($\alpha_{graphene}$) can be obtained from Eq. (7).

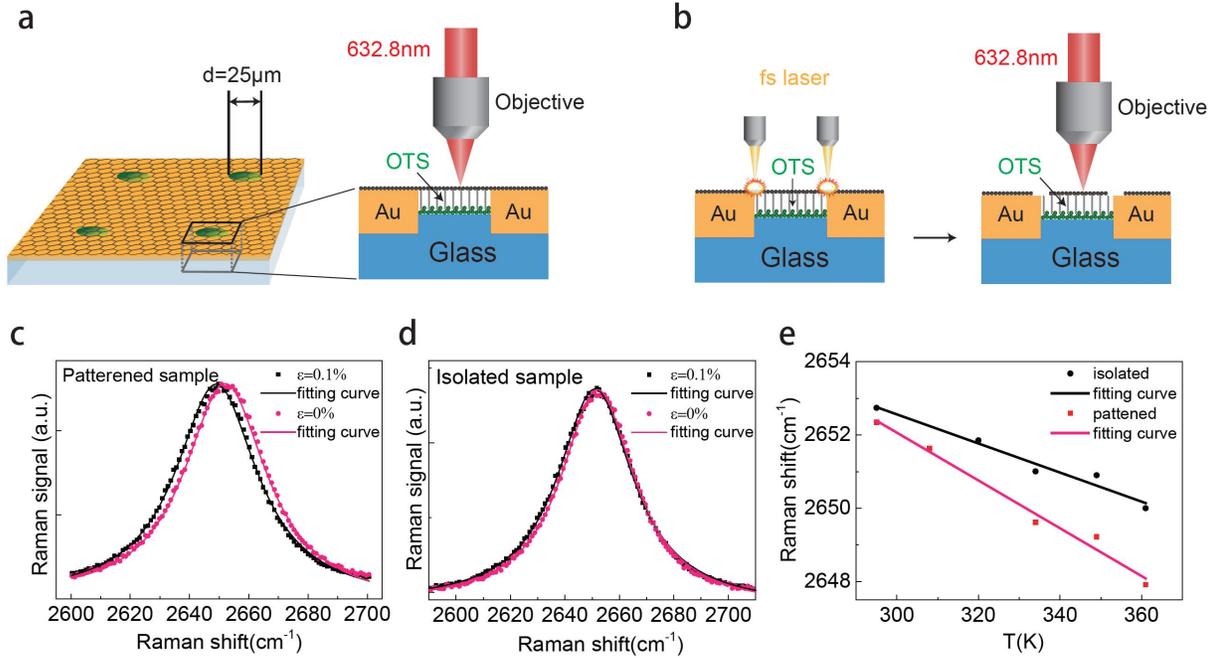

**Figure 4** (a) Sketch of graphene on patterned substrate. The green disks represent the domains with OTS monolayer on glass. The gold film (marked in yellow) is 30 nm in thickness on glass. (b) Graphene on OTS is isolated from that on gold via etching graphene along OTS edge by tightly focused femtosecond laser. Raman spectra of graphene on OTS domain of the patterned substrate under different biaxial strain before (c) and after (d) the etching. (e) Raman frequency shift as a function of temperature for the patterned (black dots) and isolated (red squares) graphene on OTS. The red and black lines are linear fitting.

Figures 4(c) and 4(d) show the Raman spectra of the 2D band for the patterned and isolated graphene on OTS, respectively. The $β_{eff}$ is found to be 38cm$^{-1}$/% and 9cm$^{-1}$/%, respectively. The latter is nearly the same as that of graphene on OTS without patterning shown in Fig. 2(a). This indicates the isolated graphene on OTS disk is decoupled from the surrounding graphene on gold film. Figure 4(e) presents the Raman shift as a function of temperature before and after laser cutting of the graphene pattern. The obvious slope difference before (black line) and after (red) laser cutting can be observed. Inserting the values of $\Delta\omega_{ls}(T_S)$, $\alpha_{\text{sub}}$ (8.0×10$^{-6}$/K, see supporting information for details) and $\beta_{\text{eff}}$ in Eq.(7), we obtained $\alpha_{\text{graphene}}$ = (-0.6±0.5)×10$^{-6}$/K, which is in agreement with the diminishing TEC from molecular dynamic simulation described previously. Our result gives a much smaller TEC of graphene at room temperature than previous reported data.[17] The difference should be attributed to an inaccurate strain coefficient ($\beta_{\text{eff}}$) and more importantly the neglect of the out-of-plane coupling between graphene and substrate in early studies.

In summary, we have studied the thermal properties of monolayer graphene supported on OTS-coated glass using in situ Raman micro-spectroscopy and molecular dynamic simulations. We showed clearly that the out-of-plane coupling between graphene and OTS can fundamentally change the thermal expansion coefficient, although their interaction is rather weak. Theoretical analysis found that the reduction of the density of states for longer wavelength out-of-plane vibrational modes is responsible for the suppression of TEC when graphene is supported on a substrate. The modeling on a perfectly flat single layer substrate diminished the negative TEC to the extent of being positive as a result of out-of-plane coupling with the substrate. The experimental OTS substrate has more irregularity, thus the TEC shows substantial reduction without sign change. Nonetheless the pronounced TEC reduction with out-of-plane coupling is clear from both theory and experiment. It alerts that in order to deduce TEC of a 2D material on substrate, besides the lateral strain, the out-of-plane coupling effect is of critical importance and must be taken into consideration in experiment, which is unfortunately ignored in literature. The TEC measurement scheme proposed in this work offers a solution that is able to separate the contributions of lateral strain and out-of-plane coupling to TEC. Our results provide new insight on thermal property and substrate effects of 2D material, which will help the development of future applications like next-generation electronic and optoelectronic devices.

AUTHOR INFORMATION

**Corresponding Author**

* Corresponding electronic mails: fengwang@uark.edu, cstian@fudan.edu.cn

**Author Contributions**

The manuscript was written through contributions of all authors. All authors have given approval to the final version of the manuscript. ‡ Q.C.F, D.S.W and Y.D.S contributed equally to this work.

**Funding Sources**

CST acknowledges support from the National Natural Science Foundation of China (11874123) and the National Key Research and Development Program of China (2016YFA0300902). FW acknowledges support from the National Science Foundation DMR-1609650.

For Table of Contents Only

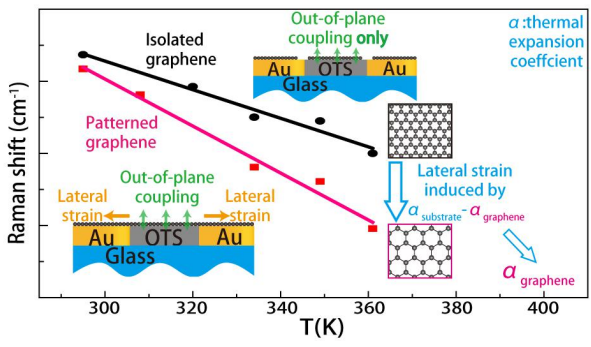